# Model performance indicators ERP systems

setare yaghubi
student, Department of computer
Zanjan Branch, Islamic Azad University
Zanjan, Iran
setareyaghubi@yahoo.com

Nasser modiri
Assoc. Prof, Department of computer
Zanjan Branch, Islamic Azad University
Zanjan, Iran
nassermodiri@yahoo.com

Masoud Rafighi
PHD Student, Department of computer
Qom University
Qom, Iran
Masoud_r62@yahoo.com

*Abstract*- Implementation process ERP is complex and expensive process. Typically always be faced with many failures. Successfully implemented in an organization has many challenges. Organizations in the deployment and success of the system depends on several factors.One of the key factors in the successful deployment of systems methodology is the implementation process. Methodology has several indicators for successful implementation of ERP systems, we have examined. And indicators for each of the methodologies have identified. The proposed method is also an important indicator of the success of security controls and indicators to be monitored and controlled.

## I. Introduction

In today's dynamic and unpredictable business environment, companies face the tremendous challenge of expanding markets and rising customer expectations. This compels them to lower total costs in the entire supply chain, shorten throughput times, reduce inventories, expand product choice, provide more reliable delivery dates and better customer service, improve quality, and efficiently coordinate globe demand, supply and production [1].

The acronym ERP (Enterprise Resource Planning) was first used in 1990 by firm Gartner Group in connection to MRP (Material Requirements Planning, later Manufacturing Resource Planning) and CIM (Computer-Integrated Manufacturing) and since then it has become widely used for streamlining not only manufacturing process but all other business processes. ERP packages integrate solutions for different directions such as accounting, contracts, payroll, maintenance and human resources management, attempting to provide technology solutions for all core functions of an enterprise, regardless of its specific business flow, such as non-manufacturing businesses, non-profit organizations and governments [2].

Deploying an ERP project are including very complexity and risks. Deployment that would painlessly to replace all current software systems of Customer and all users do not prevent new systems, if not impossible, is very difficult. Because it is faced with very problems such as don't ready information, user countering with change, impossible convert operation and some case reenter information. For countering with these problems need to deployment methodology. The methodology should be an important indicator of the success of ERP systems in organizations.

One of the main points that exist in implementing large information systems, using a methodology developed and implemented the system. This is very important in case of ERP systems due to the special nature and the impress all the processes and activities of the organization.

Section 2 in this paper we study important factors in the successful deployment of ERP systems. Section 3 introduces the indicator of the methods existed in deployment . The proposed measures are presented in Section 4 , and finally conclusions are content .

## II. critical success factors of ERP systems

Implementing an ERP system is not an inexpensive or risk-free venture. In fact, 65% of executives believe that ERP systems have at least a moderate chance of hurting their businesses because of the potential for implementation problems .It is therefore worthwhile to examine the factors that, to a great extent, determine whether the implementation will be successful. Numerous authors have identified a variety of factors that can be considered to be critical to the success of an ERP implementation [3].

Based on empirical studies of ERP implementations in organizations for successful implementation of integrated ERP systems, project management, educating and culturing, setting project goals and scope, conforming with organizational processes, software configuration, project team, development and defender of project methodologies and etc are important factors. The following are explained some of the factors successful.

A. Top Management Support/Commitment

Top management support was consistently identified as the most important and crucial success factor in ERP system implementation projects (Welti, 1999; Davenport, 1998a; Sumner, 1999; Bingi, et al., 1999; Gupta, 2000; Bancroft, et al., 1998).

Welti (1999) suggested that active top management is important to provide enough resources, fast decisions, and support the acceptance of the project throughout the company. Jarrar, et al. (2000) pointed out that the top management support and commitment does not end with initiation and facilitation, but must extend to the full implementation of an ERP system [4].

B. Define clear goals and objectives

ERP implementations require that key people throughout the organization create a clear, compelling vision of how the company should operate in order to satisfy customers, empower employees, and facilitate suppliers for the next three to five years. There must also be clear definitions of goals, expectations, and deliverables. Finally, the organization must carefully define why the ERP system is being implemented and what critical business needs the system will address [3].

C. Business Plan and Vision

A clear business plan and vision should be behind the implementation strategy to know in which direction the project must be steered. In project management three often competing and interrelated goals that need to be met are mentioned: scope, time, and cost goals. There must be a clear business plan how the goals can be achieved.

Business plan and vision summarises the CSFs clear goals and objectives, management (of) expectations, anticipated benefits from ERP implementation project, business plan and vision, adequate ERP implementation strategy, motivation behind ERP implementation, multi site issues and business case [5].

D. Project Team

ERP implementation teams should be composed of top-notch people who are chosen for their skills, past accomplishments, reputation, and flexibility. These people should be entrusted with critical decision making responsibility .Management should constantly communicate with the team, but should also enable empowered, rapid decision making .

The implementation team is important because it is responsible for creating the initial, detailed project plan or overall schedule for the entire project, assigning responsibilities for various activities and determining due dates. The team also makes sure that all necessary resources will beavailable as needed [3].

E. Communication

As the goal of ERP systems is to integrate various business functions across different locations, interdepartmental cooperation and communication is the core of the ERP implementation process (Akkermans and Helden, 2002), he suggested that intensive communication between the key parties is directly linked to the success of the project [6].

### III. Methods of deployment ERP systems

One of the main points that exist in implementation of large information systems, utilizing an approach and methodology of development and implementation of system. The importance of this in ERP system is the nature of ERP systems and impacted all processes and activities of the organization.

Methodologies are crucial for the deployment of ERP systems can refereed AIM method that is created by Orale company, ASAP methodology by SAP company and Signature by Epicor Scala company. The large firm's producer using a specific methodology for the deployment of ERP systems. ASAP methodology has five phases that is a comprehensive and rich approach and, significantly reducing the overall cost and quality of the work is done at a high level [7]. In this method there are support from project management, member of team, external consultants and technical consultants, business process [7]. and a great tool for small and medium businesses [8]. Project management is a critical factor in the implementation of ERP systems, is provided by the ASAP. Good project management, especially in the process of designing, testing and end user training are important factors in successful implementation by SAP in the most organizations [6]. ASAP is a fast and flexible methods [7,8].

Methodology provided by the company Oracle is named Application Implementation Method (AIM). This methodology involves defining the activities, work processes, standards, procedures and practices, that detail of it described in six different sections with Milestone set guide and valuation activities relative to each other and define the main activities for the speedy implementation of projects and activities are complementary choice. In order to successfully implement this methodology, first action is required, and the resources needed to do and the resource needed for accomplish a specific project are recognized, and secondly, to do all of the activities, provides a patterns for the outputs . The main advantage of this methodology is that business requirements are defined early in the project and during implementation Consider placed. One of the major disadvantages of this methodology, its complexity [7]. Framework that are including elements such as steps, processes and tasks. AIM has a very wide scope, in this field investment of firms, sectors and there is a group of branches [9].

Signature methodology used in small and medium businesses [10]. Reducing the overall cost of the system and the attitude same all of Scala consultants in all areas project.

From aspect of learning provides set of standard Classroom and Training Web-based.

A. Methodology ASAP

Figure 1 shows the phases in ASAP methodology

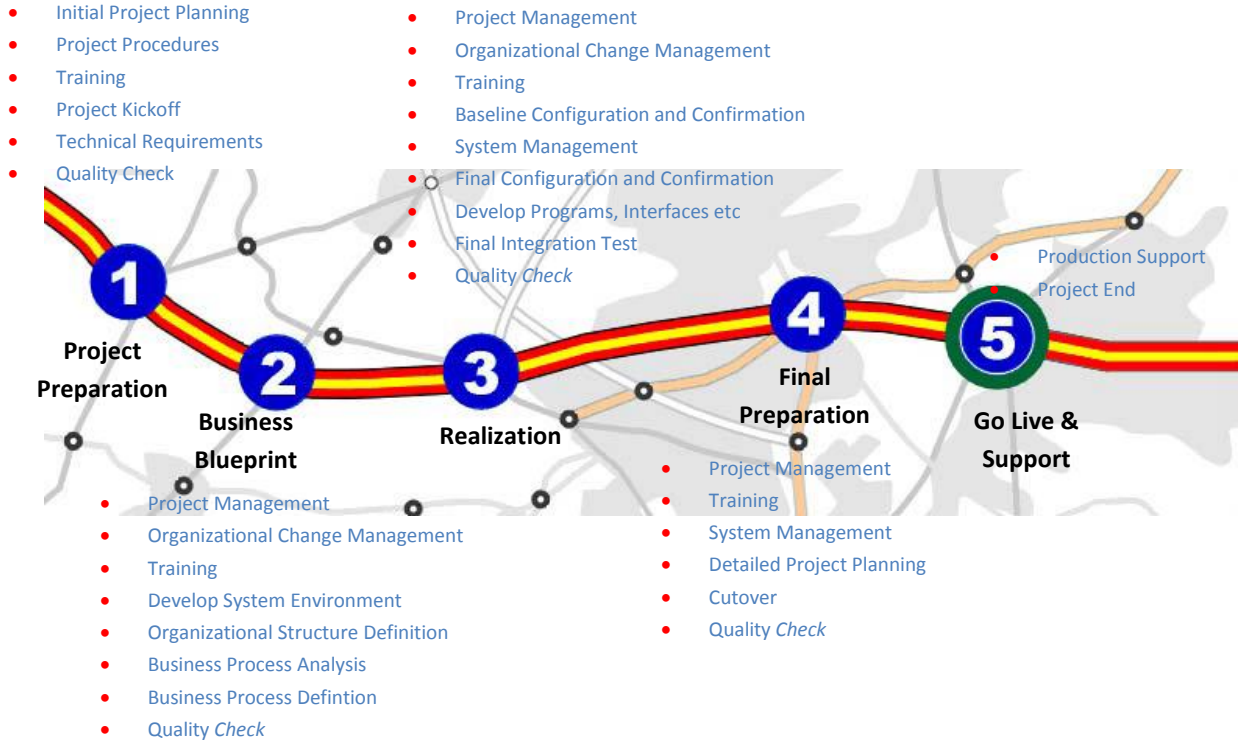

- Initial Project Planning
- Project Procedures
- Training
- Project Kickoff
- Technical Requirements
- Quality Check

**1 Project Preparation**

- Project Management
- Organizational Change Management
- Training
- Develop System Environment
- Organizational Structure Definition
- Business Process Analysis
- Business Process Defintion
- Quality *Check*

**2 Business Blueprint**

- Project Management
- Organizational Change Management
- Training
- Baseline Configuration and Confirmation
- System Management
- Final Configuration and Confirmation
- Develop Programs, Interfaces etc
- Final Integration Test
- Quality *Check*

**3 Realization**

- Project Management
- Training
- System Management
- Detailed Project Planning
- Cutover
- Quality *Check*

**4 Final Preparation**

- Production Support
- Project End

**5 Go Live & Support**

Figure 1: phases in ASAP methodology [12]

Phase I - Project Preparation

In this phase is done the preparation of elementary tasks of project [11]. Also in this step preparation of project charter, create Structure and organization of project review and refine the plan and implementing strategy implementation solutions, creating work teams and assign tasks, create plan and detailed action plan , defining technical requirements , initial meetings for bing project identification process model for doing , modeling and analysis of project requirements and determine organization of project , etc  Is performed [12].

Phase II - Business BluePrint

Review, identify and design business processes in different fields is done via a standard procedure at this stage. In additional review and modeling of the business objectives and existing structures that are done [13]. In this phase after modeling current processes of organization, achieve to step modeling the future state of organization after changes [8].

Phase III – Realization

In this phase, tasks such Training project team, initial configuration of the system and receive confirmation, changes in basic ERP software has become an appropriate ERP solution to customize by means defined, creating and testing necessary interfaces, creating reporting Tools and testing them, testing integrity of the system is performed. Also in this phase program for transition is regulated [8,12].

Phase IV - Final Preparation

This phase allocated final preparation and review the plans projects. In this phase, system administration and user training, final testing of the system, applies the modifications and changes, transfer data from old systems to new systems is performed [8,12].

Phase V: Go Live & Support

Preparation and review launching System, correcting errors, preparing plans and schedules of timing and supporting its and activities of closing project in the final phase will be performed [8,12].

B. Methodology AIM

Figure 2 shows the phases and operations of the AIM methodology

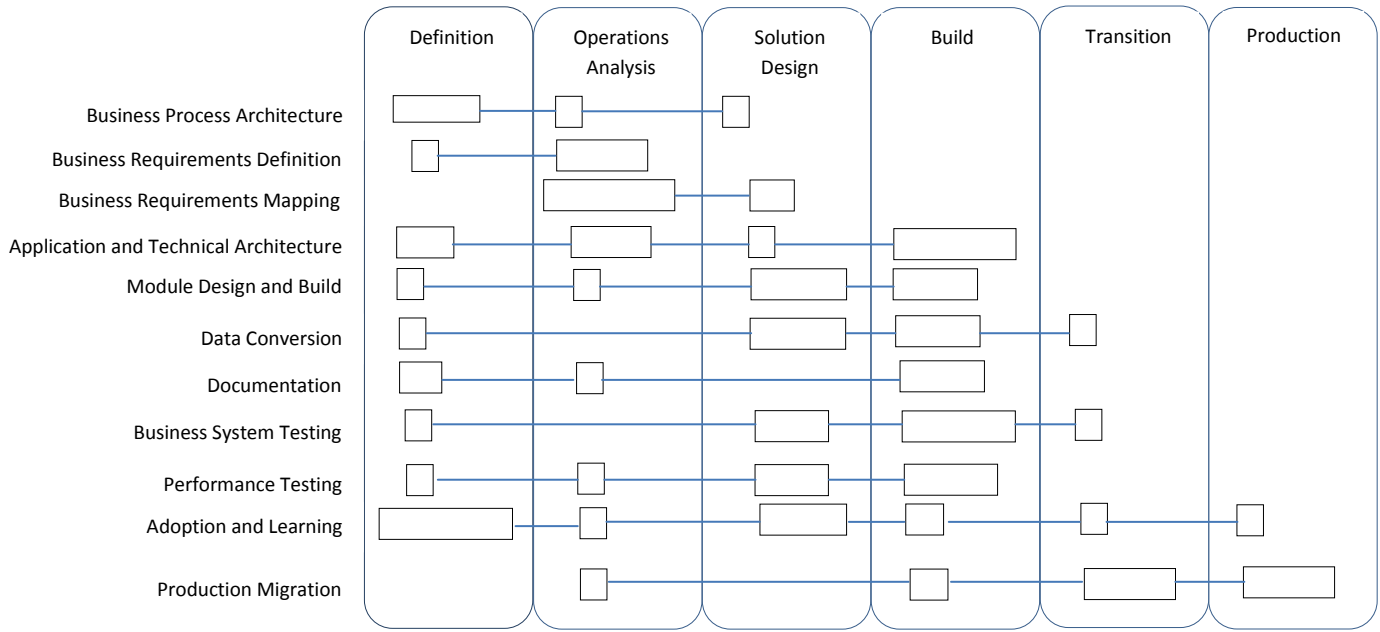

Figure 2: phases of the AIM methodology [9]

Phase I: Definition

In this phase, the employer and Contractor working together for planning process, review the resources and constraints, project scope and organization of operational teams [9,14].

Phase II: Operations Analysis

In this phase, the project team will move towards collecting work processes which can be extract differents current work step with standard applications ERP. Also, decide on the future organization of work processes are performed at this stage [9,14].

Phase III: Solution Design

The purpose of this phase detailed design new solution for the business needs and organization. Also based on organization's needs and if it is optional, it can be added selection other features can be added in solution. In this phase based on decisions of phase II provided solutions for conforming current work processes with standard process ERP [9,14].

Phase IV: Build

Coding and testing all areas of custom software, conforming software application in the organization, conversion data and design user interfaces is formed in this phase. effective testing and System testing is done in this phase. The purpose of this phase is to formulate and provide detailed requirements for computer applications and present a solution [9,14].

Phase V: Transition

At this stage programs are made in previous phase the implemented operationally in organization and the data in the system before being transferred to the new systems and its weaknesses, is amended as. In this phase, the current business processes and ERP applications are working parallel. In other words application programs are tested in a real environment [9,14].

Phase VI: Production

The final delivery of the new system at last phase of this methodology and the beginning of system support cycle has been done. Improvement and steps of measurement to be carried out at this stage [9,14].

C. Methodology Signature

Figure 3 shows the phases in the methodology Signature.

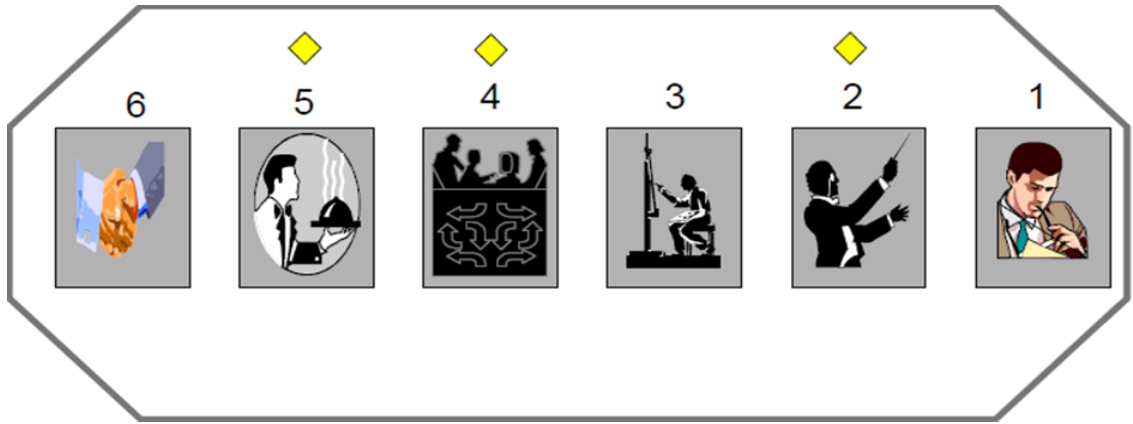

Figure 3: phases of methodology Signature.

Phase I: Analysis

The objectives of this phase definition of client's business requirements, identify key individuals of the project team, and the design of the prototype system. This phase will evaluate business requirements, key members of the project team trained [15].

Phase II: Project Organization

At this stage, the project objectives are set and approved, project plan preparation, proofing and approval, the project team and its schedule is determined. Project management and control procedures are in phase [15].

Phase III- Design

The purpose of this phase is to design and configure the system. Design business strategies, design inputs, outputs and intermediate users set the parameters of the test system and training end-users in all phases of the system are achieved [15].

Phase IV - Data Preparation

This phase of the procurement, transfer, and data validation has been allocated. Activities in this phase include: Define data conversion requirements, and construction and determine conversion methods, data convert to the new system and ensure the integrity of transmitted data [15].

Phase V - Test Run

The purpose of this phase is to review the project to ensure the proper functioning of the system. In addition, the pilot implementation of the system according to the requirements of business and the final configuration phase takes place in the system [15].

Phase VI – Hand Over

The final phase is allocated to operating new system, closure and delivery of projects. Operating system, project evaluation, quality control, delivery to support team will do well in this phase [15].

There are important indicators in each of methodologies presented. in the major indices are for the implementation of ERP systems can refereed to planned and project management, business process re-engineering, training and test and quality control.

Methodologies to deployment an integrated ERP system is an key success factor. And there are indicators in each of these methodologies that this is one of the success factors. For example, if the planning index is not, The project will most likely fail. Although the methods are deployed with the same parameters But the difference in one or more indicators is caused better success in the implementation of projects and minimizes the risk of project failure.

Project on Management is one of the most important indicators. This includes change and communication management and systems management and etc. Good project management in the process of designing, testing, and training, etc, one of the important factors is for successful

ERP implementation in the organization. Table 1 lists the most important indicators of the ways it shows.

| | Index | AIM | ASAP | Signature |
|---|---|---|---|---|
| 1 | programming | ✓ | ✓ | ✓ |
| 2 | Fast implementation | ✓ | ✓ | |
| 3 | Customization | ✓ | ✓ | ✓ |
| 4 | Project management | ✓ | ✓ | ✓ |
| 5 | Business Goals | ✓ | ✓ | ✓ |
| 6 | Risk reduction | ✓ | ✓ | |
| 7 | Support for complex projects | ✓ | | |
| 8 | Education | ✓ | ✓ | ✓ |
| 9 | Architecture SOA | ✓ | ✓ | ✓ |
| 10 | Test | ✓ | ✓ | ✓ |
| 11 | Reengineering | ✓ | ✓ | ✓ |
| 12 | Configuration | ✓ | ✓ | ✓ |
| 13 | Repeatability | ✓ | ✓ | |
| 14 | Preparation | ✓ | ✓ | ✓ |
| 15 | Business more efficient | ✓ | ✓ | ✓ |
| 16 | Flexibility | ✓ | ✓ | |
| 17 | Time reduction | ✓ | ✓ | |
| 18 | Cost reduction | ✓ | ✓ | ✓ |
| 19 | Complete coverage of the entire project life cycle | | ✓ | |
| 20 | Designing TO-Be | | ✓ | |
| 21 | Designing As-Is | ✓ | | ✓ |
| 22 | Implement a comprehensive and powerful | | ✓ | |
| 23 | Clear definition of project scope | ✓ | ✓ | |
| 24 | Incremental Model | ✓ | ✓ | |
| 25 | Support | ✓ | ✓ | ✓ |
| 26 | Help Desk | | ✓ | |
| 27 | Documentation | ✓ | ✓ | ✓ |
| 28 | Multi-site | ✓ | ✓ | ✓ |
| 29 | Multi nationality - linguistic | ✓ | | |

Table 1: important indicators of ERP

### IV. Proposed Method

In section 3 for the implementation of ERP systems indexes methodologies are studied and evaluated. In this regard, after reviewing the existing indicators, In the methodologies to deployment we recommended indicators for successful implementation of the systems ERP.

Security Index - Since ERP systems with resources and information on the organization and sometimes outside agencies interact in the matter and determine the boundaries of information security important to access this information. As a result, information security, and set limits for access to this information is important. Since the different layers vary in different organizations and the type and amount of information they are different. when implementing ERP systems in the organization should be a good strategy to control access to systems and data should be considered . Security software is a software quality assurance activity that focuses on the detection of potential risks and may have a negative impact on software as well as ruining the entire system, under three headings to integrate security, privacy and accessibility is divided. Due to that none of the methods proposed methodology is no way to control security. The proposed method, called security control measures in the will proposed methodology. That methodology will be presented in the strategy and planning and management security controls. Simultaneously with the planning and management of security controls to control costs.

Indicators of Quality Control - Quality control is a set of open projections, review and test the process used software product in order to ensure compliance with the requirements. For which it is defined. Quality control includes feedback loops, a process that has created a working product. Quality control parameters for planning, management, methods used for quality control and testing to detect errors in the different stages. Table 2 shows the parameters in the proposed method.

| | Index | AIM | ASAP | Signature | proposed method |
|---|---|---|---|---|---|
| 1 | programming | ✓ | ✓ | ✓ | ✓ |
| 2 | Fast implementation | ✓ | ✓ | ✓ | ✓ |
| 3 | Customization | ✓ | ✓ | ✓ | ✓ |
| 4 | Project management | ✓ | ✓ | ✓ | ✓ |
| 5 | Business Goals | ✓ | ✓ | ✓ | ✓ |
| 6 | Risk reduction | ✓ | | | |
| 7 | Support for complex projects | ✓ | ✓ | ✓ | ✓ |
| 8 | Education | ✓ | ✓ | ✓ | ✓ |
| 9 | Architecture SOA | ✓ | ✓ | ✓ | ✓ |
| 10 | Test | ✓ | ✓ | ✓ | ✓ |
| 11 | Reengineering | ✓ | ✓ | ✓ | ✓ |
| 12 | Configuration | ✓ | ✓ | ✓ | ✓ |
| 13 | Repeatability | ✓ | ✓ | | ✓ |
| 14 | Preparation | ✓ | ✓ | ✓ | ✓ |
| 15 | Business more efficient | ✓ | ✓ | ✓ | ✓ |
| 16 | Flexibility | ✓ | ✓ | | ✓ |
| 17 | Time reduction | ✓ | ✓ | | ✓ |
| 18 | Cost reduction | ✓ | ✓ | ✓ | ✓ |
| 19 | Complete coverage of the entire project life cycle | | ✓ | | ✓ |
| 20 | Designing TO-Be | | ✓ | | ✓ |
| 21 | Designing As-Is | ✓ | | ✓ | |
| 22 | Implement a comprehensive and powerful | | ✓ | | ✓ |
| 23 | Clear definition of project scope | ✓ | ✓ | | ✓ |
| 24 | Incremental Model | ✓ | ✓ | | ✓ |
| 25 | Support | ✓ | ✓ | ✓ | ✓ |
| 26 | Help Desk | | ✓ | | |
| 27 | Documentation | ✓ | ✓ | ✓ | ✓ |
| 28 | Multi-site | ✓ | ✓ | ✓ | ✓ |
| 29 | Multi nationality - linguistic | ✓ | | | ✓ |
| 30 | Quality control | | | | ✓ |
| 31 | Security | | | | ✓ |
| 32 | Supervision | | | | ✓ |

Table 2. List of parameters in the proposed method

## V. Conclusion

ERP is a software solution to integrate enterprise resources is beneficial. In this regard that the goal of software engineering to produce quality products in accordance with cost control is considered, therefore, to achieve this requires strategy and planning and management methodology that processes for quality control in should be considered. And implementation of a framework for the process to be used, Security software is a software quality assurance activities, in this regard that the methodologies is an important factor in the deployment, so if the security controls and control procedures should be implemented, increases the likelihood of successful implementation of ERP systems in organizations.